\documentclass[draft,jgrga]{agutex}

  \usepackage{graphicx}
\usepackage{amssymb,amsmath} 
\usepackage{color}
  \setkeys{Gin}{draft=false}
\authorrunninghead{ZHAO ET. AL.}

\titlerunninghead{GCR MODULATION DURING SOLAR MINIMUM 23/24}

\authoraddr{L.-L. Zhao,
State Key Laboratory of Space Weather, 
National Space Science Center, Chinese Academy of Sciences,
P.O. Box 8701, Beijing 100190, China,
(llzhao@spaceweather.ac.cn)}

\authoraddr{G. Qin,
State Key Laboratory of Space Weather, 
National Space Science Center, Chinese Academy of Sciences,
P.O. Box 8701, Beijing 100190, China,
(gqin@spaceweather.ac.cn)}

\authoraddr{M. Zhang,
Department of Physics and Space Science, Florida Institute of Technology, Melbourne, 
FL 32901, USA., (mzhang@fit.edu)}

\authoraddr{B. Heber,
Institut f\"{u}r Experimentelle und Angewandte Physik, 
Christian-Albrechts-Universit\"{a}t zu Kiel, 
24118 Kiel, Germany, (heber@physik.uni-kiel.de)}

\begin{document}

%
%

\title{Modulation of galactic cosmic rays during the unusual solar minimum between
cycles 23 and 24}
%

%
%



\authors{L.-L. Zhao\altaffilmark{1}, G. Qin\altaffilmark{1}, 
M. Zhang\altaffilmark{2}, and B. Heber\altaffilmark{3}}

\altaffiltext{1}{State Key Laboratory of Space Weather, National Space Science Center, Chinese Academy of Sciences, Beijing 100190,
     China}
\altaffiltext{2}{Department of Physics and Space Science, Florida Institute of
     Technology, Melbourne, FL 32901, USA}
\altaffiltext{3}{Institut f\"{u}r Experimentelle und Angewandte Physik, Christian-Albrechts-Universit\"{a}t zu Kiel, 
24118 Kiel, Germany}



%
%


\begin{abstract}
During the recent solar minimum between cycles 23 and 24 (solar minimum 
$P_{23/24}$) the intensity of Galactic 
Cosmic Rays (GCRs) measured at the Earth was the highest ever recorded since space 
age.
It is the purpose of this paper to resolve the most plausible mechanism 
for this unusually high intensity. A GCR transport model in three-dimensional 
heliosphere based on a simulation of Markov stochastic process is used to find the
relation of cosmic ray modulation to various 
transport parameters, including solar wind (SW) speed, 
distance of heliospheric boundary, magnitude of interplanetary magnetic field (IMF) 
at the Earth, tilt angle of heliospheric current sheet (HCS), 
values of parallel and perpendicular diffusion coefficients.
We calculate GCR proton energy spectra at 
the Earth for the last three solar minima $P_{21/22}$, $P_{22/23}$, and $P_{23/24}$,
with the transport parameters obtained from observations. 
Besides weak IMF magnitude and slow SW speed, we find that a possible 
low magnetic turbulence, which increases the 
parallel diffusion and reduces the perpendicular diffusion in the polar direction, 
might be an additional possible mechanism for the high GCR intensity in the solar 
minimum $P_{23/24}$.
\end{abstract}

%
%

%

\begin{article}

%
%

\section{Introduction}
Galactic Cosmic Rays (GCRs) are energetic charged particles originated far away from 
the heliosphere. The high energy GCRs may reach the Earth atmosphere to produce 
secondary elementary particles that can be measured by ground-based Neutron Monitors (NMs)
or other detectors. 
Although the lower energy GCRs (tens of MeV/nuc) are not usually detected by the 
ground-based NMs, they can be measured in space by spacecraft except during solar energetic
particle (SEP) events produced by solar flares or coronal mass ejections. 
Unlike SEPs, GCRs form a nearly stable and isotropic background of high-energy radiation. 
The intensity of GCRs is slowly modulated in an
anti-correlation \citep{McDonald98} with the solar activity level of 11-year cycle. 
It occurs because GCR particles have to travel through the magnetized interplanetary
medium. The interplanetary magnetic field emanated from the Sun
changes with the solar cycle, causing variations in the speed of particle
transport processes such as diffusion, 
convection, adiabatic deceleration and drifts. Therefore, 
GCRs can provide important information about their propagation
and modulation mechanisms in  the heliosphere \citep{kot12}. Once the level of modulation
is figured out, we can reconstruct the spectrum and composition of GCRs
in the interstellar space, which can further provide information about their origin and
the acceleration mechanism that produces them at the source.

\begin{figure}[htbp]
\centering
\includegraphics[width=\textwidth]{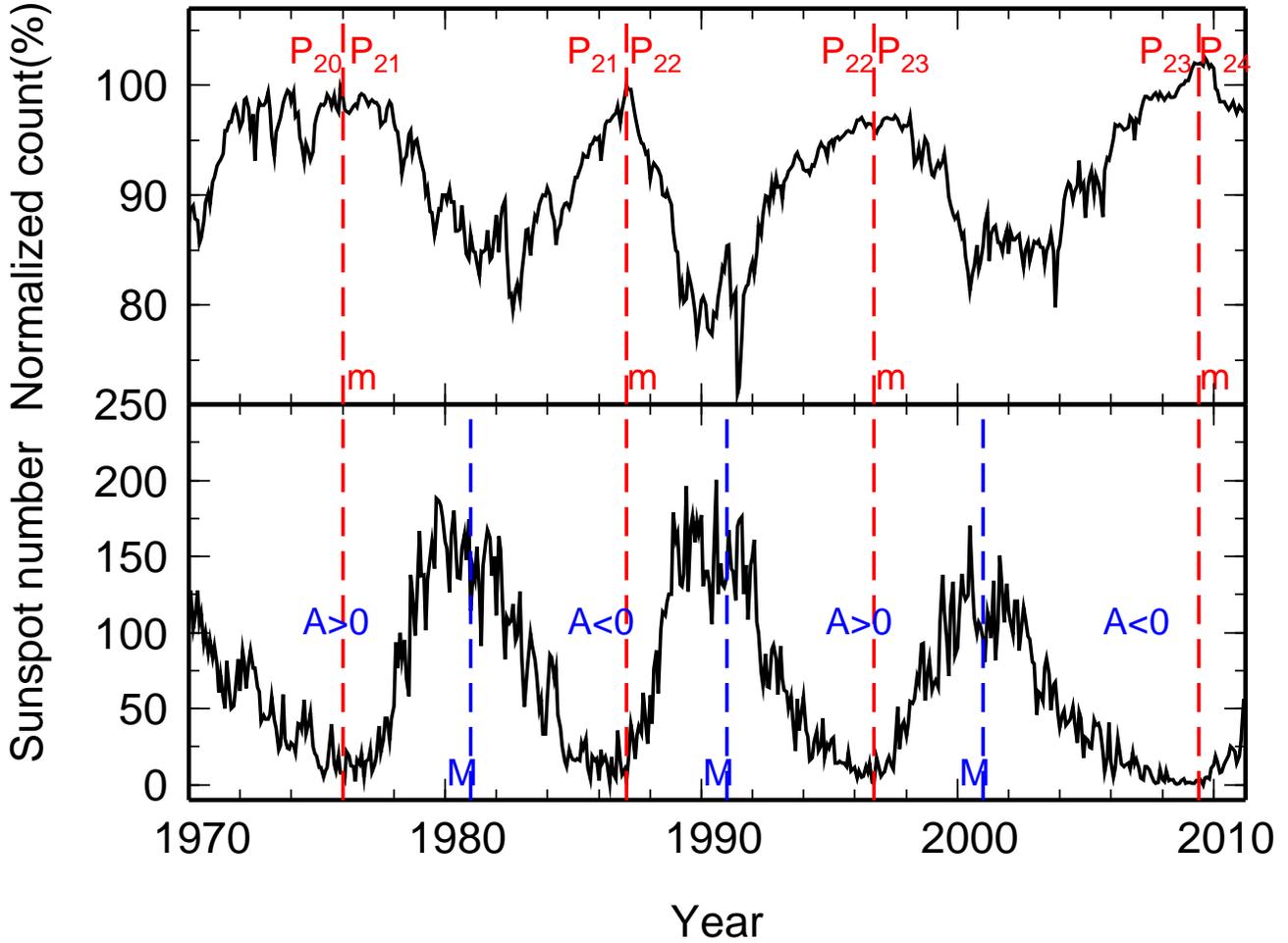}
\caption{GCR intensity as measured by Apatity NM (upper panel) and monthly averaged 
SSN (lower panel). The red dashed-lines indicate the epochs 
of solar minima, and the red numbers represent solar cycles. The black dashed-lines 
indicate the epochs of solar maxima, and `$A > 0$' or `$A < 0$' represent the 
periods of solar magnetic polarity.}\label{fig:sunspot}
\end{figure}

The GCRs intensity measured at the Earth reached a record high level
during the last solar minimum 
between cycles 23 and 24, noted as solar minimum $P_{23/24}$ from now on. 
Figure \ref{fig:sunspot} shows the GCR count rates as measured by the
Apatity NM, whose effective cutoff rigidity is 0.65 GV,
 and the monthly averaged SunSpot Numbers (SSNs) for the past forty years.  
The red dashed-lines indicate the epochs of solar minima, which demarcate the solar 
cycles represented by the red numbers from the next ones. The black 
dashed-lines indicate the epochs of solar maxima, which demarcate the periods of 
solar magnetic polarity represented by `$A>0$' or `$A<0$'. From 
Figure \ref{fig:sunspot} we can clearly see a few well-known features of GCRs. First, 
an anti-correlation between GCR intensity and 11-year solar activity cycles is 
shown. Second, in the cycles with $A<0$ magnetic polarity like 1980s, and 2000s, when the 
Interplanetary Magnetic Field (IMF) points towards (outwards) the Sun in the 
northern (southern) hemisphere \citep{sch04}, the time profiles of positively 
charged particles in the GCR are peaked, 
whereas the time profile is more or less flat in the cycle of $A>0$ 
magnetic polarity like 1970s and 1990s. This phenomenon is attributed to the 
``waviness'' of the Heliospheric Current Sheet (HCS) \citep[see][]{kot83}. Besides 
the above characteristic
 behavior, we can also notice that the monthly mean SSN reached a minimum 
value around 2009. It was followed by a high GCR count rate which breaks the previous
record  February 1987 level. Meanwhile, the Solar Wind (SW) density, pressure, 
and IMF strength all reached the lowest values ever observed during the latest measurements 
made by Ulysses \citep{heb09}. 

Various models, empirical and theoretical \citep[e.g.,][]{ahl10, man11}, have been used
to study the unusual GCRs intensities during this solar minimum. The empirical 
and phenomenological GCRs modulation models are derived from observations without 
considering the physical processes \citep[e.g.,][]{nymmik92,zhao12}. But in order to
understand the physical causes for such phenomenon, one needs to use theoretical 
models for GCR modulation. The most successful ones are based on 
\citet{par65}, which essentially includes all important GCR modulation mechanisms such as outward 
convection by the SW, diffusion through the irregular IMF, gradient and curvature 
drifts, adiabatic deceleration from the divergence of the expanding SW. 
\citet{bur89} further concluded that GCR drift in the tilted HCS can be an important effect 
in solar modulation of GCR. The variation of particle perpendicular diffusion through 
the changes in magnetic field turbulence may also cause different levels of modulation. 
Recent studies also show that there is remarkable modulation in the outer heliosphere \citep{Scherer2011APJ}, probably as well as beyond the heliopause \citep{Strauss2013APJL, Strauss2014ASR}.
Therefore, 
the GCR intensities measured at Earth is a comprehensive result of these 
different conditions for particle propagation through the heliosphere. 
More detailed theories were summarized
in review papers such as \citet{pot98}, \citet{jok00}, \citet{heb06} and \citet{pot13}. 
 Finite-difference method \citep{jok79, kot83} and stochastic method 
\citep{zha99, bal05, pei10} have been used to solve the 2-D or 3-D
Parker's transport equation for GCR modulation. Calculation results
were able to reproduce many observed features from measurements by spacecraft, balloon experiments, and NMs. 
Although the study of GCR modulation has been progressed significantly, 
much work still need to be done. The record level of GCR intensity during
the last solar minimum naturally throw us a question: what causes the unusual
solar minimum?

It is the purpose of this paper to answer the question of what causes the unusually high GCR
intensity at Earth in the last solar minimum. 
We first present 
the observations of SW and IMF parameters measured at $1$ AU for the last several solar 
cycles. Next we use a GCR transport model with numerical simulation to study the 
modulation of cosmic rays. Finally, through comparing our simulation results with 
the observations, we show what are the possible reasons for the unusual high GCR intensity for the
last solar minimum $P_{23/24}$. 

\section{Modulation Model}

The distribution function of cosmic rays propagating through the
heliosphere is governed by Parker
transport equation \citep{par65},
\begin{equation}
\frac{\partial f}{\partial t} = \nabla \cdot(\kappa\cdot\nabla f)-(
   \mathbf{V}_{sw}+\mathbf{V}_d)\cdot \nabla f + \frac{p}{3}(\nabla \cdot 
   \mathbf{V}_{sw}) \frac{\partial f}{\partial p} ,
  \label{eq:parker}
\end{equation}
where $f(\mathbf{r},p)$ is the cosmic ray distribution function, with
$p$ the particle's momentum, $\mathbf{r}$ the particle's position, $\mathbf{V}_{sw}$
the SW speed, and $\mathbf{V}_d$ the gradient and curvature drifts in the 
IMF. 
The spatial diffusion coefficient tensor $\kappa$ is diagonal, and consists of a 
parallel diffusion coefficient $\kappa_{\parallel}$ and two perpendicular diffusion 
coefficients, $\kappa_{\perp r}$ the perpendicular diffusion coefficient in the 
radial direction and $\kappa_{\perp \theta}$ that in the polar direction. Here we 
assume the parameters are axially symmetric and time-independent on the time scale of
average particle transport through the heliosphere as discussed below. In 
addition, we assume the IMF as a Parker spiral, and that the SW velocity is radial 
from the sun and constant in magnitude. 
Note that cosmic ray is considered 
isotropic, otherwise the adiabatic deceleration term, the last one in the right hand 
side of equation (\ref{eq:parker}), has to be in the anisotropic form 
\citep[e.g.,][]{QinEA04}.

In this work a relatively simple spatial and momentum dependence of the diffusion 
coefficients is assumed following \citet{zha99} and \citet{fer01}. Firstly,
parallel diffusion is set as \citep{zha99,fer01} 
\begin{equation}
\kappa_{\parallel}=d\kappa_0\beta\left(\frac{p}{p_0}\right)^{\gamma}
\left(\frac{B_e}{B}\right)^\eta,
\label{eq:kappaparallel}
\end{equation}
with the parallel diffusion factor $d$ being an adjustable constant, 
$\kappa_0=1\times10^{22}$ cm${}^2$ s${}^{-1}$, $\gamma=1/3$, 
$\eta=1$, $\beta$ is a fraction of particle's speed relative to the 
speed of light, $p_0=1$ GeV c${}^{-1}$ is a reference
momentum, $B_e$ is the magnetic field strength at the Earth, and $B$ is the magnetic
field at the location of the particle. Note that we set $\gamma=1/3$ according to 
QLT of cosmic rays \citep{jok66} for a Kolmogorov turbulence spectrum. 
However, other parameter from a Kraichnans scaling could also be used.
Note that the form of diffusion coefficient for cosmic ray propagation
 in the heliosphere is rather complicated \citep[e.g.,][]{MatthaeusEA03,qin07,ShalchiEA04,ZankEA04}. 
 For example, it is assumed that a break in the rigidity-dependent parallel diffusion 
coefficient around $4$ GV is necessary for explaining the observed boron-to-carbon 
ratio \citep{bus08, sha10}.
In this work we use diffusion forms without break for the simplicity purpose. 
Since the peak of GCR spectrum at solar minimum is well below $1$ GeV and the level of modulation is much lower for $> 4$ GV GCR, the effect of the break on modulated spectrum is insignificant. 
Secondly, the diffusion coefficients in the two perpendicular directions are set to proportional to
the parallel diffusion coefficient according to test particle simulations
\citep[e.g.,][]{gia99, qin02, qin07},
\begin{equation}
\kappa_{\perp r} = a\kappa_\parallel/d=a\kappa_0\beta\left(\frac{p}
{p_0}\right)^{\gamma}\left(\frac{B_e}{B}\right)^\eta
\label{eq:kappar},
\end{equation}
with an adjustable constant factor $a$ for the radial perpendicular diffusion, and
\begin{equation}
\kappa_{\perp \theta} =b\kappa_\parallel/d=b\kappa_0\beta\left(\frac{p}
{p_0}\right)^{\gamma}\left(\frac{B_e}{B}\right)^\eta
\label{eq:kappatheta},
\end{equation}
with an adjustable constant  factor $b$ for the polar diffusion perpendicular diffusion.
Here, we assume different values of the parameters $a$ and $b$ for
non-axisymmetric perpendicular diffusion because of non-axisymmetry of turbulence 
\citep[e.g.,][]{MatthaeusEA03} or the background magnetic field. Note that 
\citet{eff12apj}
also discussed the effects of different perpendicular diffusion coefficients.

We also include a wavy HCS provided by \citet{jok81}, who showed that if the solar
wind velocity is radial and constant in magnitude, the HCS can be represented by 
\begin{equation}   
\theta^\prime=\frac{\pi}{2}+\sin^{-1}\left[\sin\alpha\sin\left(\phi-\phi_0+
\frac{r\Omega}{V_{sw}}\right)\right],
\end{equation}
where $\alpha$ is the HCS tilt angle (TA), $\phi_0$ is an arbitrary azimuthal phase constant, and $\Omega$ is 
the angular velocity of the Sun's rotation corresponding to a period of 27.27 days.
Furthermore, if the TA $\alpha \ll 1$, the HCS can be approximately written as
\begin{equation}
\theta^\prime\approx \frac{\pi}{2}+\alpha\sin\left(\phi-\phi_0+\frac{r\Omega}{V_{sw}}
\right).
\label{eq:HCS}
\end{equation}
Next, using the approximate form of HCS equation (\ref{eq:HCS}) we can express the Parker's spiral IMF as,
\begin{equation}
\mathbf{B}=\frac{A}{r^2}\left(\mathbf{\hat e}_r-\frac{r\Omega\sin{\theta}}{V_{sw}}
\mathbf{\hat e}_{\phi}\right)\left[1-2H \left(\theta-\theta^{'}\right)\right],
\label{eq:IMF}
\end{equation}
where $A$ is used to determine the strength and polarity of IMF, 
with pointing either outward ($A>0$) or inward 
($A<0$) in the northern hemisphere. The Heaviside step function $H$ is used to 
switch the field's direction across the HCS at $\theta=\theta^\prime$.
Note that a Fisk field with latitude-dependent solar wind speed should
be used in 3D modeling,  but \citet{HitgeABurger10} found that the solar wind speed does not
significantly influence cosmic ray transport in most conditions. Therefore, for the simplicity 
purpose, here we use Parker field with constant solar wind speed.

We describe drifts in the IMF in two different ways following \citet{bur89}. 
Particles whose gyro motion do not cross the HCS 
have a pitch-angle averaged drift velocity given by 
the guiding center approximation. 
Derived with equation (\ref{eq:IMF}), the 
regular drift velocity of a particle with charge $q$, momentum $p$, and speed $v$ can be written as
\begin{eqnarray}
\mathbf{V}_{dr} &=& \frac{pv}{3q}\mathbf{\nabla \times}\left(\frac{\mathbf{B}}{B^2}\right) \nonumber \\
  &=& \frac{2pvr}{3qA(1+\Gamma^2)^2}
  \left[-\frac{\Gamma}{\tan\theta} \mathbf{\hat e}_r + (2+
\Gamma^2)\Gamma\mathbf{\hat e}_\theta +\frac{\Gamma^2}{\tan\theta}
\mathbf{\hat e}_\phi\right], 
\label{eq:Vdr}
\end{eqnarray}
where $\Gamma=r\Omega\sin\theta/V_{sw}$ is the tangent of the angle between the direction of IMF and the radial direction  $\mathbf{\hat e}_r$. 
Particles with a trajectory that crosses the HCS will experience a fast meandering drift along the HCS. 
Assuming a locally flat HCS, the magnitude of the drift velocity $v_{ns}$ along the HCS can be approximated as \citep[see also][]{bur89}
\begin{equation}
  v_{ns}=\left\{0.457-0.412\frac{d}{r_L}+0.0915\left(\frac{d}{r_L}\right)^2\right\}v,  \quad \quad \quad \quad for \quad |d| < 2 r_L   
\label{eq:Vdcs}
\end{equation}
{\color{black}
where $d$ is the distance from the position of the particle to the HCS, $r_L$ is gyroradius, and $v$ is the particle speed.
{\color{black}Calculation results with this realistic HCS drift is the 
same as those with analytical HCS drift of \citet{kot83}.}
The direction of the HCS drift velocity is parallel to the HCS and perpendicular to the HMF (\cite[e.g.][]{bur89}). See \cite{bur12} for detailed discussion on the drift velocity direction in 3-D HCS.
}
{\color{black}Note that both the drift expressions (equation (\ref{eq:Vdr}) and (\ref{eq:Vdcs})) are only valid when scattering is neglected,
{\color{black}which is the case for solar minimum.}}

The inner boundary is set at {\color{black}$r=0.3$ AU} as an absorption boundary.
The outer boundary of the heliosphere, which assumed as {\color{black}the heliopause (HP) at $r=R_{\text{HP}}$}, is set to be a GCR source with {\color{black}an assumed local interstellar spectrum (LIS)}
\begin{equation}
  J_{LIS} \propto p(m_0^2c^2+p^2)^{-1.8}
\label{eq:fp}
\end{equation}
by following \citet{zha99}. 
Though it is believed that
with measurements from Voyager 1 spacecraft in the vicinity of the heliopause \citep{Decker2012Nature} and highly accurate measurements by the PAMELA mission \citep{Adriani2011Science}, it is now possible to determine the lower limit of the very LIS for protons, helium and other ions with numerical simulations \citep{Herbst2012APJ}.
Nevertheless, the true LIS is still far from conclusive \citep{Webber2013}.
In addition, different LIS models can produce the observed 
spectrum with LIS model-dependent modulation parameters \citep{herbst10}. 
Furthermore, recent studies show that remarkable modulation exists in the outer heliosphere
and even beyond the heliopause \citep[e.g.][]{Scherer2011APJ, Strauss2013APJL}.
And the outer heliospheric structure and boundary of the dynamic heliosphere also change associated with the varying solar activity \citep{Zank2003JGR, Scherer2003GRL, Pogorelov2009SSR}.
However, assuming a steady LIS during the studied period, a distance of the 
boundary, and an inclusion of the heliosheath just has minor effects for modulation 
at $1$ AU, since most of the energy loss occurs in the inner heliosphere.
Here we study the modualtion process within the inner heliosphere, 
so only the LIS without other effects over the boundary is considered for simplicity purpose.


\section{Interplanetary Environment} 
 
\begin{figure}[htbp]
\centering
\includegraphics[width=\textwidth]{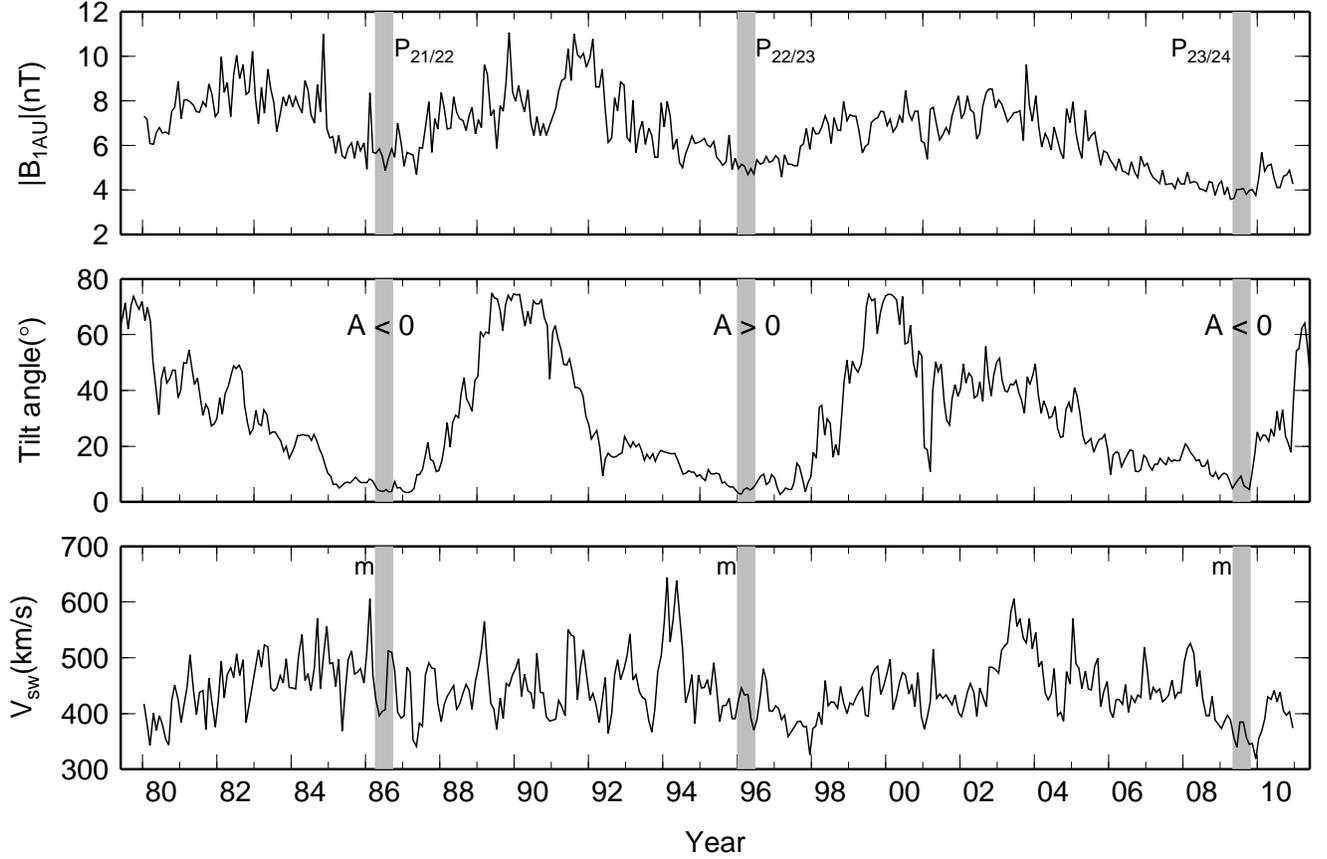}
\caption{Temporal evolution interplanetary solar wind and magnetic field 
parameters measured at $1$ AU. The IMF 
and SW speed are obtained by averaging the OMNI data over one-month intervals. The TA 
of the HCS is obtained from the WSO Web site with ``new'' model.  The three grey 
shadow areas labeled with $P_{21/22}$, $P_{22/23}$ and $P_{23/24}$ indicate the three (21/22, 22/23, 
and 23/24) epochs of the solar minimum of approximately half a year long.}
\label{fig:cbvt}
\end{figure}

In order to understand  solar modulation of GCR with model simulations using the transport equation (\ref{eq:parker}), 
it is important to use appropriate particle transport parameters, which
are determined by the properties of the solar wind, heliospheric magnetic field, and
energetic particles.  Figure \ref{fig:cbvt} 
shows the temporal evolution of IMF $B_e$ and SW speed $V_{sw}$, both of which are
measured at $1$ AU, and the HCS TA $\alpha$, for the last three 
solar cycles. The IMF and the SW velocity data are obtained by averaging the OMNI 
data over one-month intervals. And the TA of the HCS data are obtained from the WSO 
Web site with the ``new'' model. In Figure \ref{fig:cbvt}, we illustrate the three 
epochs of solar minima in grey shadows of about half a year long as $P_{21/22}$ 
(1986, 91 - 1986, 273), $P_{22/23}$ (1996, 1 - 1996, 182), and $P_{23/24}$ 
(2009, 121 - 2009, 304). Note that all the data during the solar minima in this work are
averaged over the periods shown above.
From Figure \ref{fig:cbvt} we can see that both the magnitude of IMF and the SW 
speed are very low during the recent solar minimum $P_{23/24}$, but the TA of HCS is
 not at the lowest level. 

The solar magnetic polarity and the half year average of $V_{sw}$, $B_e$, and 
$\alpha$ during the three solar minima, which are used in our simulations for GCR 
modulation, are shown in the Table \ref{tbl-pms}.

\section{Numerical Methods} \label{method}
There are many approximate solutions of the Parker equation available, e.g. the most generally used force field solution \citep{Moraal2013SSR}.
The appeal of the force field approach lies in the fact that 
observed modulation can be described with a single parameter termed modulation potential $\phi$ \citep{Lopez2004JGR}.
The model assume an equilibrium between diffusion and adiabatic energy loss. Effects of
drift and convection are neglected.
While it is possible to reproduce the observed GCR modulation in the inner heliosphere through adjusting the modulation potential $\phi$ using the force field model, it cannot resolve the contribution from distinct physical mechanisms.

In this work, we use the time-backward Markov stochastic process method proposed by \citet{zha99}
to solve the Parker transport equation in 3-D spherical coordinate (\ref{eq:parker}). 
As it is more versatile and less
 computationally intensive, this method has been successfully implemented with different cosmic
ray transport models, such as \citet{qin05} and \citet{bal05}. In this method,
we trace virtual particles from the observation point back to the outer boundary with
 the interstellar flux expressed  as equation (\ref{eq:fp}).
Note that the GCR protons distribution is written as $j\sim p^2f$. 
{\color{black}The set of SDEs, being equivalent to equation (\ref{eq:parker}), for a pseudo-particle in position ($r$, $\theta$, $\phi$) and momentum $p$ using spherical coordinate can be written as equation (\ref{eq:ito}) \citep[see also][]{pei10,stra12}.
\cite{kopp12} and \cite{eff12} also present a general discussion on the SDE technique for solving Parker transport equation.} 
{\color{black}
\begin{eqnarray}
  \mathrm{d}r & = & \left[\frac{1}{r^2}\frac{\partial}{\partial r}(r^2 \kappa_{rr})+\frac{1}{r\sin\theta}\frac{\partial\kappa_{r\phi}}{\partial\phi}-V_{sw}-v_{dr}\right]\mathrm{d}s \nonumber \\
  & & +\sqrt{2\kappa_{rr}-\frac{2\kappa^2_{r\phi}}{\kappa_{\phi\phi}}}\mathrm{d}W_r+\frac{\sqrt{2}\kappa_{r\phi}}{\sqrt{\kappa_{\phi\phi}}}\mathrm{d}W_\phi, \nonumber\\
  \mathrm{d}\theta &=&\left[\frac{1}{r^2\sin\theta}\frac{\partial}{\partial\theta}(\sin\theta\kappa_{\theta\theta})-\frac{v_{d\theta}}{r}\right]\mathrm{d}s+\frac{\sqrt{2\kappa_{\theta\theta}}}{r}\mathrm{d}W_\theta, \\
  \mathrm{d}\phi & =& \left[\frac{1}{r^2\sin^2\theta}\frac{\kappa_{\phi\phi}}{\partial\phi}+\frac{1}{r^2\sin\theta}\frac{\partial}{\partial r}(r\kappa_{r\phi})-\frac{v_{d\phi}}{r\sin\theta}\right]\mathrm{d}s \nonumber\\
  & & +\frac{\sqrt{2\kappa_{\phi\phi}}}{r\sin\theta}\mathrm{d}W_\phi, \nonumber\\
  \mathrm{d}p & = & \frac{p}{3r^2}\frac{\partial r^2 V_{sw}}{\partial r}\mathrm{d}s. \nonumber
\label{eq:ito}
\end{eqnarray}
}
Using the stochastic simulation, we can obtain not only modulated GCRs fluxes, but also the behavior of individual particle, e.g. the propagation time and energy loss \citep{Strauss2011JGR}.
In addition, we can incorporate almost any kind of 
magnetic field configuration according to observations or MHD numerical simulations \citep{Strauss2013APJL}.
Furthermore, this stochastic numerical method is more computationally efficient than the traditional finite difference approach, with the added advantage that it is easy to parallelize.
Note that the integration of stochastic differential equation 
is performed in terms of spherical coordinates, which further {\color{blue}enhances} the computational efficiency by reducing coordinate transformations. 

\section{Modulation Effects} 

In this section the effects of various transport parameters on GCR modulation 
are discussed. Throughout this section, we set magnitude of IMF at $1$ AU $B_e=5$ 
nT, SW speed $V_{sw}=400$ km/s, TA of HCS $\alpha=0^\circ$, and heliospheric outer 
boundary distance as $80$ AU, unless otherwise stated. Note that all results 
from numerical simulations and observations are at $1$ AU in the ecliptic.

\subsection{Modulation Effects of Interplanetary Parameters}
\begin{figure}[htbp]
\centering
\includegraphics[width=\textwidth]{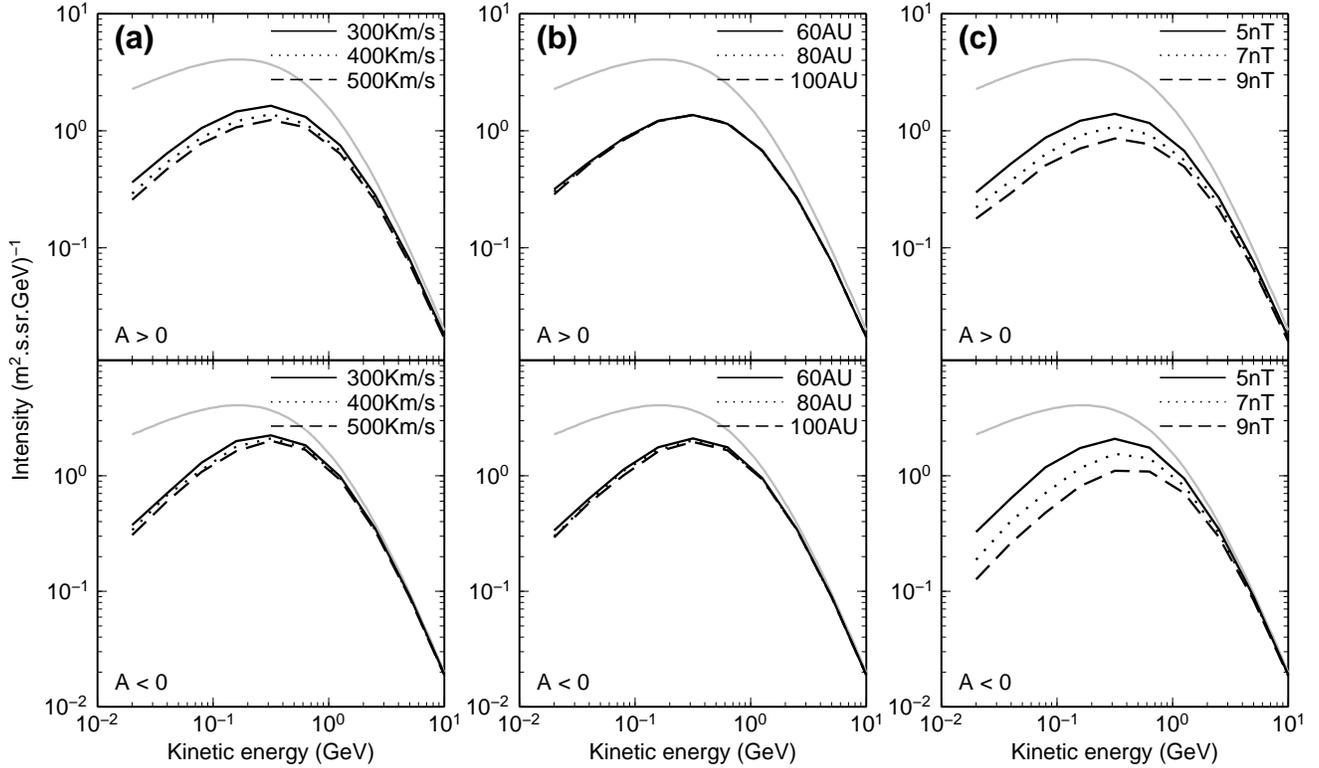}
\caption{Computed differential intensity of GCR proton at Earth as a 
function of kinetic energy for both $A>0$ and $A<0$ 
magnetic polarities during the solar minimum 
condition with an unmodulated interstellar spectrum shown in grey line as a 
reference. Three different black lines indicate three assumptions for (a) SW speed,
(b)distance of the outer heliospheric boundary, and (c) magnitude of IMF.}
\label{fig:parameter}
\end{figure}

First, we study the modulation effects of interplanetary solar wind and magnetic field parameters. 
In these simulations,  we set diffusion factors $a=0.03$, $b=0.01$ and $d=1$
in equations (\ref{eq:kappar}), (\ref{eq:kappatheta}) and (\ref{eq:kappaparallel}), 
respectively. The TA of HCS is set to $\alpha=0^\circ$ which is
appropriate for the solar minimum condition.
Figure \ref{fig:parameter} illustrates separately the computed differential intensity
for GCR protons with different interplanetary parameters used in this study.
The calculations are done for both solar magnetic polarities. 
The top panels of each figure show the results in the $A>0$ epochs, 
and the bottom panels show the results in the $A<0$ epochs,
 with the interstellar unmodulated spectrum (grey lines) for reference.

Figure \ref{fig:parameter}(a) shows the influence of different SW speeds on GCR 
proton intensity, with the dark solid, dotted, and dashed lines representing three 
assumptions of SW speed, $300$ km/s, $400$ km/s, and $500$ km/s, respectively.
Although the IMF magnitude at the Earth $B_e$ is fixed,
the magnetic field magnitude in the heliosphere is dependent on the SW speed and 
varies according to equation (\ref{eq:IMF}).
We can see there is an obvious anti-correlation between SW speed and GCR
intensity. Figure \ref{fig:parameter}(b) illustrates the influence of the heliospheric 
outer boundary radial distance on GCR intensity, with the dark solid, dotted, and
dashed lines representing three assumptions for the outer boundary radial distance, 
$60$ AU, $80$ AU, and $100$ AU, respectively. We can see that the outer boundary 
radial distance has little effect on the GCR flux measured at $1$ AU, no matter 
whether $A>0$ or $A<0$. 
In Figure \ref{fig:parameter}(c) the computed GCR proton intensities 
for different magnitude of IMF at $1$ AU are shown. 
 Compared with the results of SW speed and outer heliospheric boundary, the increased magnitude of IMF remarkably declines the GCR intensity for both magnetic epochs,
especially for the lower energy range.

Overall, Figure \ref{fig:parameter} suggests that, in our model, the low SW speed and magnitude of IMF 
play significant role in increasing the GCR flux,
 while the effect of outer heliospheric boundary is negligible.
Therefore, we set the outer heliospheric boundary distance as $80$ AU in 
the rest of the paper, but the SW speed and magnitude of IMF for each period
according to the Table \ref{tbl-pms}.

\begin{figure}[htbp]
\centering
\includegraphics[width=\textwidth]{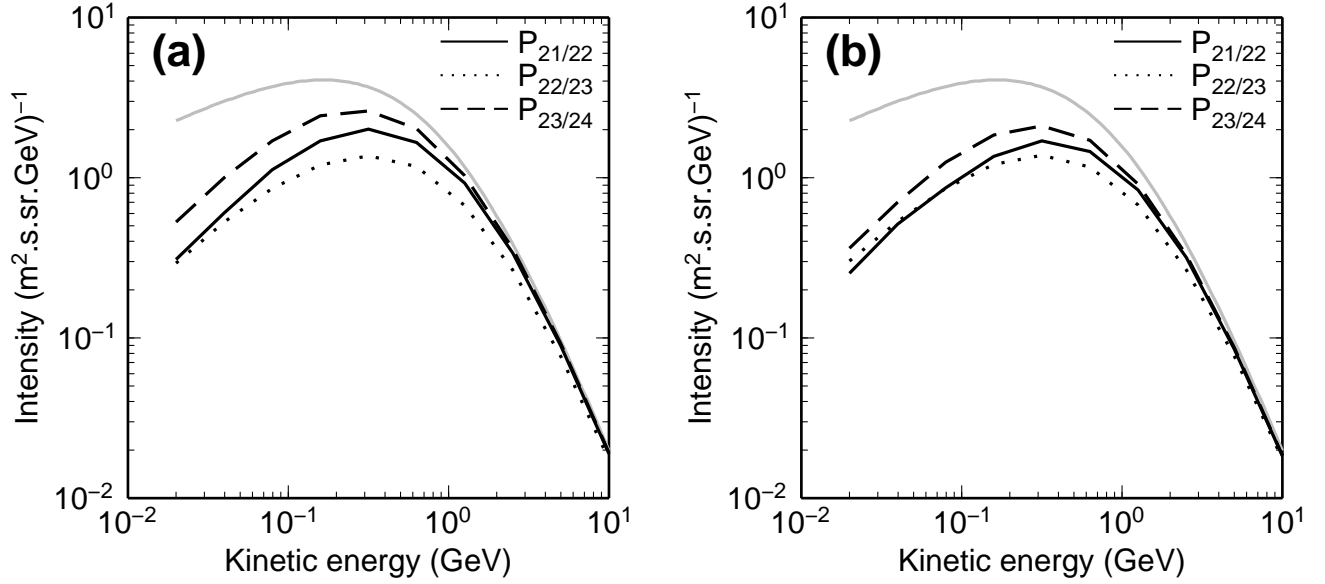}
\caption{Computed GCR proton energy spectra at the Earth for different magnetic 
field strength $B_e$ at Earth and SW speed $V_{sw}$ with unmodulated interstellar
spectrum shown in grey lines as a reference, during $P_{21/22}$ (dark solid
line), $P_{22/23}$ (dotted line), and $P_{23/24}$ (dashed line). The TA of HCS is 
set to be (a) $0^\circ$ and (b) the measured values during the corresponding periods.
}\label{fig:magnetic}
\end{figure}

In order to show the effectiveness of lower SW speed and magnitude of IMF on the
significant increase of GCR intensity in the recent extreme solar minimum, we 
calculate GCR intensities with interplanetary properties during each of the last 
three solar minima shown in Figure \ref{fig:magnetic}. Here, we set SW speed 
$V_{sw}$ and IMF magnitude at the Earth $B_e$ during the last three solar minima as
that in the Table \ref{tbl-pms}. In Figure \ref{fig:magnetic}(a), by setting TA of 
HCS as $0$, we find that the GCR intensity during $P_{23/24}$ increases 
significantly. However, in Figure \ref{fig:magnetic}(b), by setting TA of HCS for 
different solar minima as shown in Table \ref{tbl-pms}, 
the increase of the GCR intensity during $P_{23/24}$ is less prominent
compared with the spacecraft measurements shown later. 
Although the particle drifts, including the global gradient and curvature drifts, still play a significant role in CR modulation, the fact that the TA of HCS during $P_{23/24}$ is not the lowest prevent us from 
reproducing the abnormally high GCR intensity.
 So we need to consider the other physical mechanisms of modulation 
processes.

\subsection{Modulation Effects of Diffusion Coefficients}
\begin{figure}[htbp]
\centering
\includegraphics[width=\textwidth]{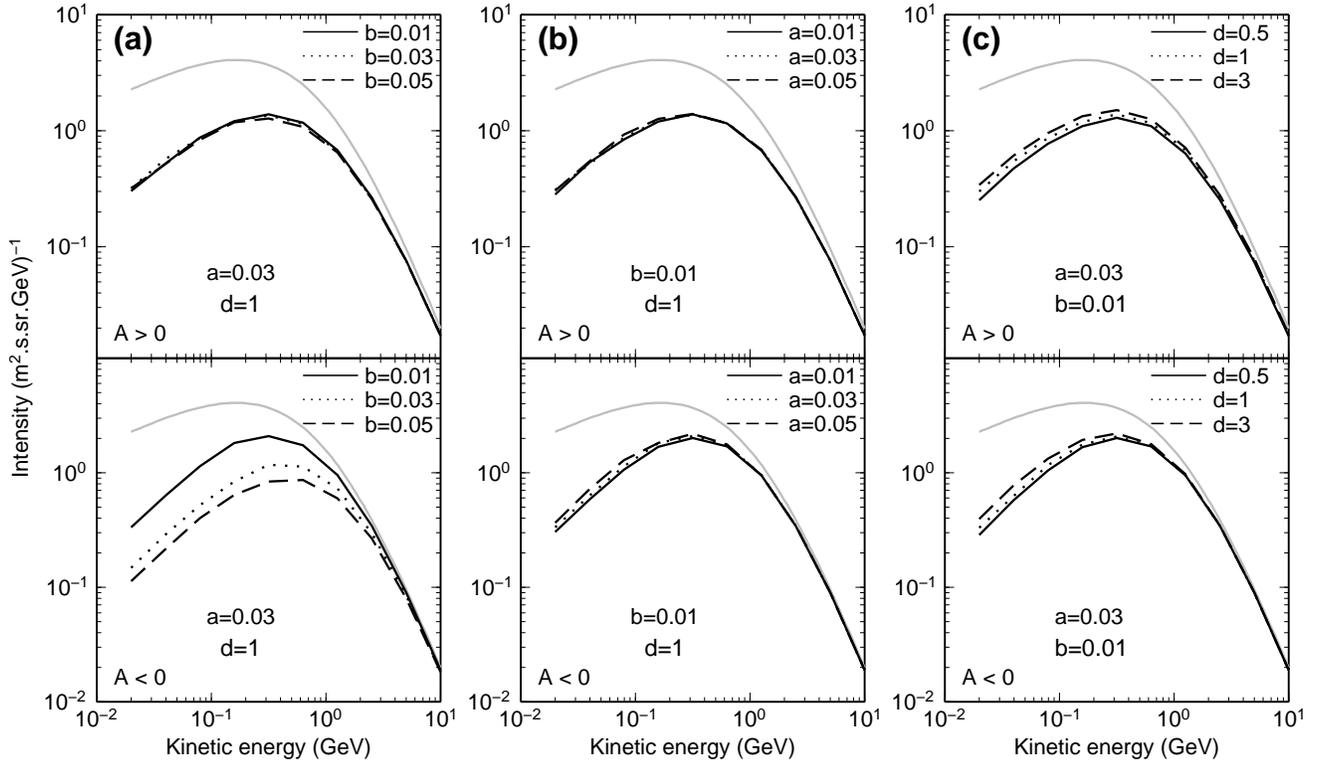}
\caption{Computed differential intensity of GCR proton at Earth as a 
function of kinetic energy for both magnetic polarities during a solar minimum 
condition with unmodulated interstellar spectrum shown in grey lines as a 
reference. Three different black lines indicate three assumptions for (a) polar perpendicular diffusion factor $b$, (b) radial perpendicular diffusion factor $a$ and (c) parallel diffusion factor $d$. 
}
\label{fig:turbulence}
\end{figure}

Since during the extreme solar minimum $P_{23/24}$, an $A<0$ epoch, the solar 
activity was unusually quiet compared to that in the other solar minima, with an 
expected lower turbulence level in solar wind, both the radial and polar 
perpendicular diffusion coefficients, $\kappa_{\perp r}$ and 
$\kappa_{\perp \theta}$, respectively, became smaller, and the parallel diffusion 
coefficient, $\kappa_\parallel$, became larger.
Here, we investigate the effects of polar perpendicular diffusion factor $b$, radial 
perpendicular diffusion factor $a$,  and  parallel diffusion factor $d$, on the GCR 
intensity, especially during an $A<0$ epoch (Figure 
\ref{fig:turbulence}).
It is similar to \citet{rein93} who discussed different
diffusion coefficients on the different intensity of CR during consecutive solar
minimum. Following \citet{eff12apj}, we also use an anisotropic diffusion
coefficients in this study.
In these simulations, we set diffusion factors $a=0.03$, $b=0.01$ and $d=1$ in equations (\ref{eq:kappar}), (\ref{eq:kappatheta}) and (\ref{eq:kappaparallel}), respectively, unless otherwise stated.

The modulation effectiveness of $\kappa_{\perp\theta}$ for both magnetic 
epochs, $A>0$ and $A<0$, is illustrated in Figure \ref{fig:turbulence}(a). Simulation 
results with $b=0.01$, $b=0.03$, and $b=0.05$ are shown with dark solid, dotted, and 
dashed lines, respectively. 
While a lower polar perpendicular diffusion factor $b$ has little effect on the GCRs 
intensity in the $A>0$ epochs, it can significantly increase GCRs intensity in 
the $A<0$ epochs. 
Moreover, Figure \ref{fig:turbulence}(b) and \ref{fig:turbulence}(c) show that higher 
radial perpendicular diffusion factor $a$ and parallel diffusion factor $d$ can 
increase GCRs intensity slightly for both solar epochs. 
Nevertheless, this effect can be significantly weakened by an decrease of polar 
perpendicular diffusion factor $b$.

The above study shows that the decrease of $b$ can cause the increase of GCR intensity.
In an $A<0$ epoch, this influence is more effective than the factor $a$ and $d$.
Therefore, it is possible to use the combined effect of these transport parameters
to explain the record level of GCR flux in $P_{23/24}$ solar minimum. Note that,
in the following simulation, the values of magnetic field magnitude $B_e$, solar wind
speed $V_{sw}$,
and tilt angle of current sheet $\alpha$ are from measurements, but the diffusion
factors $a$, $b$, and $d$ are free parameters constrained by fitting numerical
simulation results to the spacecraft measurements.

\section{GCRs Data}

In this paper, we use GCRs data from both ground based NM count rates
and proton flux of spacecraft measurements. 
The GCR data are obtained with half year average for each of three solar minimum,
$P_{21/22}$, $P_{22/23}$, and $P_{23/24}$.

The NM stations we use for GCRs data are Apatity, Oulu, Yakutsk, Moscow, 
Novosibirsk, Lomnicky Stit, Jungfraujoch, Hermanus, Rome, Tbilisi, and
Potchefstroom NMs. In order to compare GCRs count rates measured by NMs 
with flux from simulation results, we use the effective energy of each NM 
\citep{ala03}, which can be approximated as 
\begin{equation}
E_{\text{eff}}=E_1+\frac{E_2 \left(P_c/P_1\right)^{1.25}}{1+10 \exp\left(-0.45 P_c/P_1
\right)},
\label{eq:Eeff}
\end{equation}
where $P_c$ is the local geomagnetic cutoff rigidity, $E_1=6.4$ GeV, $E_2=1.45$ GeV,
and $P_1=1$ GV. Thus the integral GCR flux above the effective energy 
$M(E_{\text{eff}})$ is defined as
\begin{equation}
M(E_{\text{eff}})=\int_{E_{\text{eff}}}^{\infty}j(E)d(E)
\label{eq:integral}
\end{equation}
is directly proportional to the NM count rates, or
\begin{equation}
M(E_{\text{eff}})=K_{\text{NM}} N(P_c),
\label{eq:ME_eff}
\end{equation}
with $N(P_c)$ the NM count rates, and $K_{\text{NM}}$ a constant for any NM.
Therefore, for different NMs we can compare the computed $M(E_{\text{eff}})$ with 
observational data of the NM count rates. 
Note that the effective energy is quite different from the median 
rigidity below which lies 50\% of detector counting rate \citep{Ahluwalia2007JGR}, 
widely used for transient cosmic ray solar modulation studies \citep{Ahluwalia2013SP}. 

Table \ref{tbl-nms} shows the local 
geomagnetic cutoff rigidity $P_c$, and the corresponding effective energy $E_{\text{eff}}$ 
of NMs used in our work.

The data are obtained from STEREO and 
PAMELA for energy $22\sim 77$ MeV and $82\sim 20,000$ MeV, respectively, during the 
period $P_{23/24}$, and IMP-8 for energy $70\sim 400$ MeV during the periods 
$P_{21/22}$ and $P_{22/23}$. The data of IMP-8 and STEREO contain both GCRs and 
SEPs.
It is assumed that the modulated GCRs flux can be described as a  
stable ``background'', while SEPs appear typically as short spikes of a few days 
long except for relative higher energy particles. Therefore, similar to what was 
done in \citet{qin12} we use an automatic  despiking algorithm based on Poincar\a'e 
map thresholding method \citep{gor02} to remove the SEP spikes for STEREO and IMP-8
data. For more details to remove the SEP contamination in the time-series GCR flux 
from spacecraft observations, please refer to \citet{qin12}. 

%
%
%

\section{Simulation Results}
In the following we compare the results of our numerical simulation of GCR spectra
with measurements to find out possible reasons for the 
unusually high cosmic ray intensity during the $P_{23/24}$ solar minimum.

\begin{figure}[htbp]
\centering
\includegraphics[width=0.8\textwidth]{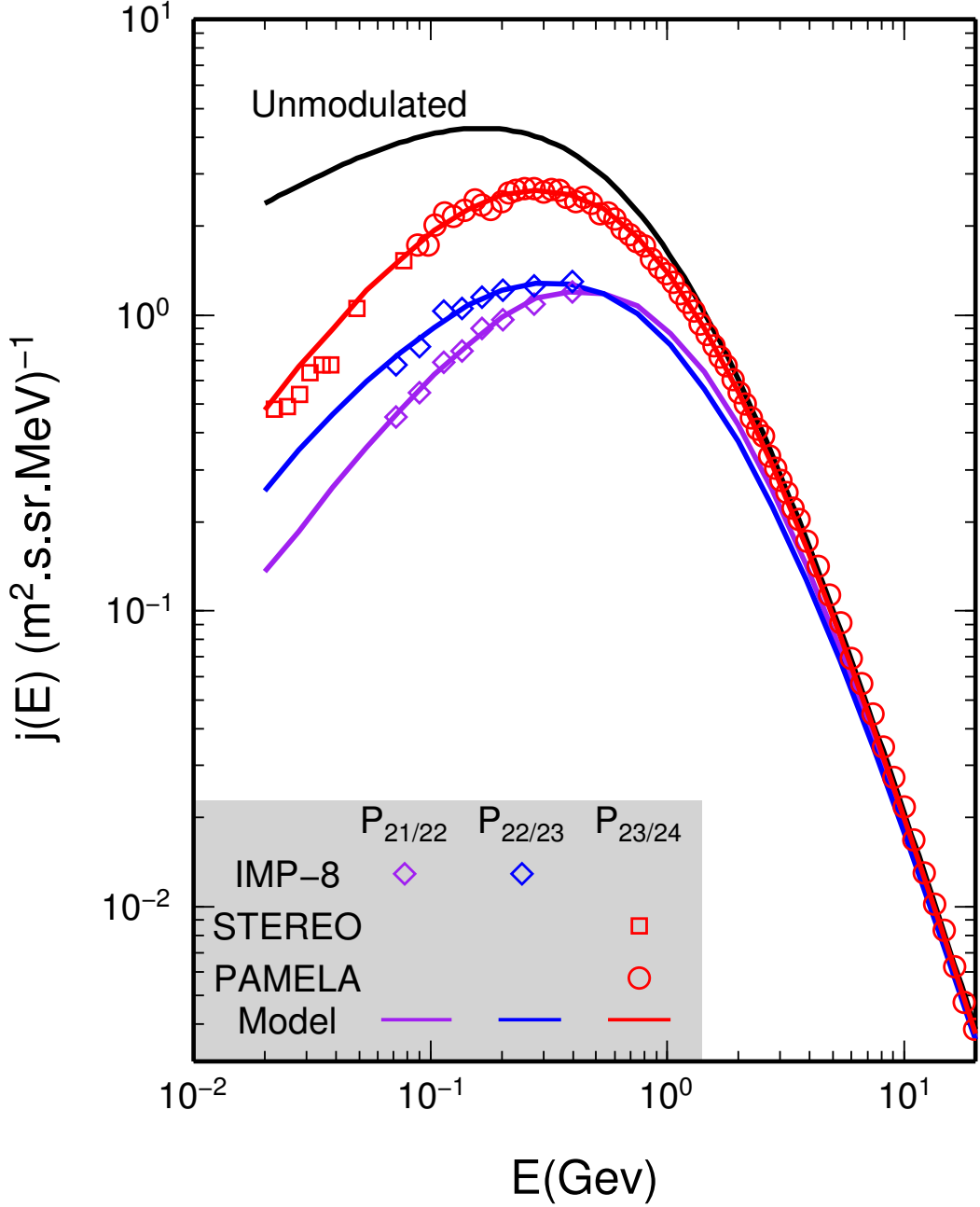}
\caption{Computed GCR proton energy spectra at the Earth for the last three solar minima with parameters presented in table \ref{tbl-pms}.
The observation data are calculated from the measurements of proton flux by STEREO (squares) and IMP-8 (diamonds) after SEP
contribution is removed.
And red circles denote the measurements from PAMELA instrument for the 
year 2009 \citep[][table 1]{adr13}.
}
\label{fig:result}
\end{figure}

Figure \ref{fig:result} shows the computed GCRs of protons energy spectra at the 
Earth for the last three solar minima with interplanetary parameters from observations 
shown in Table \ref{tbl-pms}, which include the solar magnetic polarity, magnitude of 
IMF, SW speed, TA of HCS. As a reference, black solid line indicates the unmodulated 
GCR spectrum at the outer boundary. 
Lines shown in purple, black, and red colors represent $P_{21/22}$, $P_{22/23}$, and 
$P_{23/24}$, respectively. 
IMP-8 and STEREO spacecraft measurements of 
GCRs are shown as diamonds and squares, respectively, and red circles 
denote the measurements from PAMELA instrument in the higher energy range for the 
year 2009 \citep[][Table 1]{adr13}.
For each energy point, the 
flux is calculated with a stochastic process simulation. From Figure 
\ref{fig:result} we can see that
with diffusion parameters $a=0.03$, $b=0.02$, and $d=0.5$, the simulation results 
fit well to the IMP-8 observational data during $P_{21/22}$ and $P_{22/23}$. 
As discussed earlier, in the solar minimum $P_{23/24}$ the solar activity was 
extremely quiet, so that the particles perpendicular diffusion coefficients are set 
to be smaller, and that the particles parallel diffusion coefficients are larger. 
For this reason, in $P_{23/24}$ the parameters $a$ and $b$ should be smaller and the
 parameter $d$ should be larger. From Figure \ref{fig:result} it is shown that with 
parameters $a=0.02$, $b=0.01$, and $d=1$, and other parameters set as in Table 
\ref{tbl-pms}, the simulation results fit well to the observations from both STEREO and PAMELA during the solar minimum $P_{23/24}$.

\begin{figure}[htbp]
\centering
\includegraphics[width=0.7\textwidth]{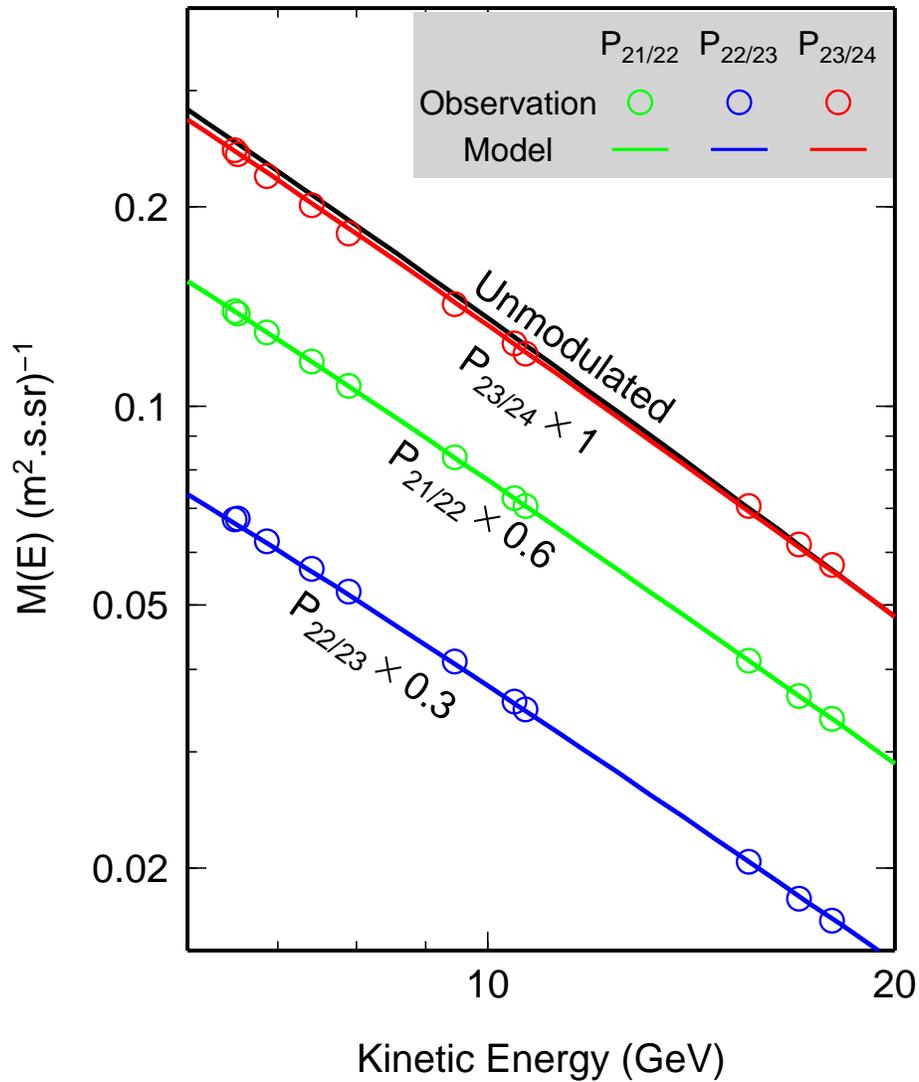}
\caption{Comparison between the computed GCR integral flux
and the NM count rates for the last three solar minima.
Note that both simulation result and observation of each solar minimum are multiplied by an arbitrary factor as denoted in figure, for the purpose of presentation.
}
\label{fig:nmresult}  
\end{figure}

Figure \ref{fig:nmresult} shows a comparison of the integral intensity $M(E)$ as a 
function of GCR energy $E$ between our simulation results and the NM measurements.  
Similar to Figure \ref{fig:result}, the black solid line indicates the unmodulated 
GCR spectrum, and the three lines in different colors represent our calculations for the three solar minima. 
Note that both simulation result and observation of each solar minimum are multiplied by an arbitrary factor for the purpose of presentation.
For each NM with a cutoff rigidity $P_c$ given in Table 
\ref{tbl-nms}, we have calculated $M(E_{\text{eff}})$ (colored lines) as an integration of 
simulated GCR flux $j(E)$ using equation (\ref{eq:integral}). In order to make a 
direct comparison between  $M(E_{\text{eff}})$ from our simulation results (green line) and
 the NM count rates in $P_{21/22}$, we obtain a normalization constant $K_{\text{NM}}$ with
 equation (\ref{eq:ME_eff}) for each NM, and we show the $K_{\text{NM}}$ in Table 
\ref{tbl-nms}. With the $K_{\text{NM}}$ we can convert all NMs' count rates $N(P_c)$ to 
their $M(E_{\text{NM}})$, which is denoted as observational data (color dots) for periods
other than $P_{21/22}$. Note that the constants $K_{\text{NM}}$ are obtained with equation 
(\ref{eq:ME_eff}) for data in $P_{21/22}$, so the green dots agree with green line 
exactly for $P_{21/22}$. For the other two solar minima, we use the same 
normalization constant and NM measurements to obtain the blue and red dots, which 
are considered as measurements. Therefore, the fact that the bule and red dots agree
 well with blue and red lines, respectively, show that our simulation results 
fit well with the NMs count rates for periods $P_{22/23}$ and $P_{23/24}$. We 
especially point out that in $P_{23/24}$, the NMs count rates were much higher than 
previous solar minima and our simulations reproduce such a phenomenon.

\begin{figure}[htbp]
\centering
\includegraphics[width=0.8\textwidth]{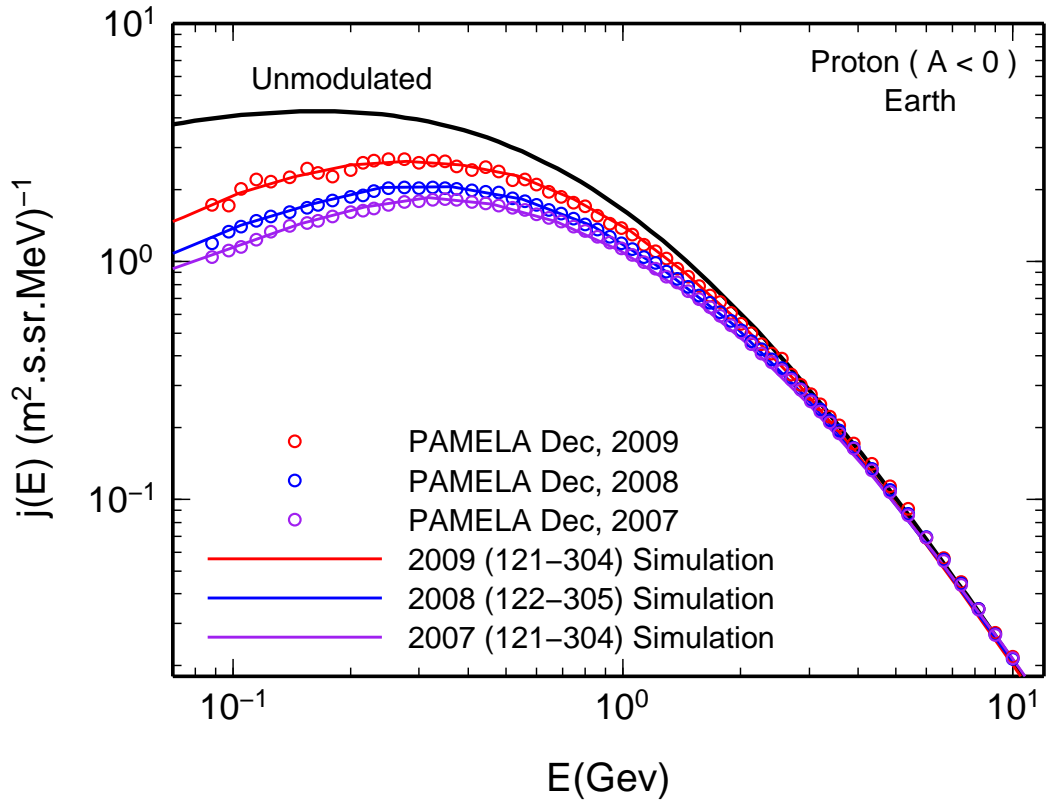}
\caption{Evolution of the proton energy spectrum during the period of minimum solar 
activity, from year 2007 to year 2009. The purple, blue and red curves indicate the 
computed GCR proton differential fluxes corresponding to three half year periods, 2007 (121-304), 2008 (122-305), and 2009 (121-304), respectively. 
An unmodulated interstellar spectrum is shown in black line for reference. 
The observations from PAMELA instrument are also shown (circles).}
\label{fig:around}
\end{figure}

Furthermore, we study the evolution of the proton energy spectrum during the period 
of the solar minimum $P_{23/24}$ (Figure \ref{fig:around}).
The proton flux measurements from PAMELA instrument with monthly average \citep[][table 1]{adr13} for year 2007, 2008 and 2009 are represented with purple, blue and red circles respectively. 
Obviously, the proton spectra in 2009 represents the highest flux observed. 
Figure \ref{fig:around} also shows the computed differential intensity of GCR protons at the Earth from 2007 to 2009 (solid lines) in half year periods. 
For simulations in these half year periods, 
the SW speed, magnitude of IMF and TA of HCS are from the averaged observations, while the diffusion coefficient parameters $a$, $b$, and $d$ are the same as the $P_{23/24}$ solar minimum.
We can see that the simulation results agree well with PAMELA measurements. 

\section{Discussion}
In this study, we investigate the behaviors of GCR modulation at Earth 
and try to determine the potential mechanisms responsible for the 
abnormally high GCR intensity in the last solar minimum, 
through comparing the numerical simulation results with the observations 
from NM stations and spacecraft instruments.

Various modulation processes could contribute to the high GCR intensity,
e.g., particle drifts, diffusion, or possible weaken outer heliosphere modulation. 
Generally, drifts effects are thought to dominant modulation process at solar minimum for $A<0$ epochs \citep{kot83}.
\cite{Cliver2011SSR} argues that diffusion is the primary modulation process during this unusual solar minimum.
\cite{Potgieter2014SP} also shows that the rigidity-dependent diffusion coefficients need to decrease significantly below $\sim3$ GeV to reproduce the proton spectra from PAMELA experiment.
In this work, we further highlight that a possible low magnetic turbulence, which increases the parallel diffusion and reduces the perpendicular diffusion in the polar direction, might be an additional mechanism for the high GCR intensity during the $P_{23/24}$ solar minimum.

Energetic particles can be scattered parallel to the background magnetic field because of 
magnetic turbulence, so higher turbulence levels would cause stronger scattering and
 shorter parallel mean free path. In addition, energetic particles perpendicular 
diffusion is achieved with the diffusive separate of particle gyrocenters caused by 
turbulence transverse complexity. Therefore, lower turbulence levels would increase 
parallel diffusion and decrease perpendicular diffusion \citep[e.g.,][]{jok66,
MatthaeusEA03,qin07}.

However, drifts still play a significant role in the modulation process, even though the 2009 solar minimum is more `diffusion dominated' than previous solar minima \citep{Potgieter2014SP}.
A low SW speed can cause less outward convection of GCRs 
out of the heliosphere and less adiabatic cooling, and a low magnitude of
IMF would cause much increase of particle drift according to equation (\ref{eq:Vdr}) 
in our model and diffusion. 
In fact, the more realistic scenario is that all modulation processes interplay dynamically, contributing to the observed increases in the proton spectra.

We can use the most advanced NLGC theory for the diffusion 
coefficients\citep[e.g., NLGC,][]{MatthaeusEA03}. The theory depends on assumption of 
turbulence type and its transport. Many free parameters are need in this theory. In 
this work, we 
assume ad hoc changes in the magnitude of diffusion coefficients. And our 
parameters only give a sense how the diffusion coefficients are expected to be.

Furthermore, the varies of magnetic turbulence properties, 
such as turbulence levels and turbulence correlation scales are important to cause the changes in diffusion coefficients. 
Nevertheless, since there is no in-situ measurement of diffusion coefficients, 
it is very difficult to estimate the diffusion coefficients quantitatively via 
comparing simulation results 
with the observations. 
Generally, there are two ways to address this issue.
On the one hand, more realistic and accurate 3-D heliospheric model should be implemented in the simulation process, e.g., solar wind profile and IMF structure from 3-D MHD numerical simulation, as well as GCR drift model within 3-D HCS structure. Moreover, modulation in the outer heliosphere should also be taken into account.
On the other hand, more observations should be used to verify this hypothesis. For instance, the electron and positron spectra observed by PAMELA provides us an unprecedented opportunity to further investigate the modulation process in this unusual solar minimum, and the high statistically significant fluxes of heavy-ions from ACE spacecraft also help to constrain the modulation model more strictly.
However, these topics are out of the scope of this paper, and we leave them for 
future study.

\section{Conclusions}

Observations of GCR count rates of NMs and the transport parameters from spacecraft 
measurements for the last three solar cycles show that during
 the solar minimum $P_{23/24}$, the intensity of GCRs was the highest, while the IMF
 and the SW speed were both weaker than the previous two solar minima, $P_{21/22}$ 
and $P_{22/23}$, but the TA of HCS was not at the lowest level. 
We first study modulation effects of the related transport parameters during the solar minimum separately, including SW speed, outer heliospheric boundary, 
magnitude of IMF at the Earth, and parallel and perpendicular diffusion 
coefficients. 
Despite the fact that drifts still play a significant role in the modulation process, we find that the particle drift during this $A < 0$ cycle cannot 
contribute solely to the high flux of GCRs in the recent solar minimum $P_{23/24}$. 
Furthermore, during the recent solar minimum $P_{23/24}$ the solar activity
was very weak and solar wind turbulence level was expected to be lower than previous
solar minima, so that particles radial and polar perpendicular diffusion 
coefficients should be smaller and parallel diffusion coefficients should be larger.
Therefore, we have to further tune the magnitude of diffusion coefficients.  It is 
found that a lower polar perpendicular diffusion with factor $b$
can cause the increase of GCRs intensity. In addition, the factor $b$ is more 
effective than the radial perpendicular diffusion factor $a$ and parallel diffusion
factor $d$ for the $A < 0$ cycle. The 
combination of lower polar diffusion coefficient, higher parallel diffusion 
coefficient, lower SW speed, and lower magnetic field in the solar minimum 
$P_{23/24}$ is possible to explain the unusually high GCR intensity. 

Although relatively simple models are implemented in our simulation model, this work represents an 
important first step towards investigating the unusual cosmic ray modulation during the last solar minimum quantitatively.
Further effort is needed to overcome these limitations in a more comprehensive way.

\begin{acknowledgements}
This work was supported in part by grants NNSFC 41125016, NNSFC 41374177, CMA grant 
GYHY201106011, and the Specialized Research Fund for State Key Laboratories of 
China. The computations were performed by Numerical Forecast Modeling R\&D 
and VR System of State Key Laboratory of Space Weather and Special HPC work stand of 
Chinese Meridian Project. We benefited from the Sunspot data provided by SIDC Team 
2009, the Wilcox Solar Observatory data obtained via the web site 
{http://wso.stanford.edua}, energetic particle data provided by IMP-8 Goddard Medium 
Energy (GME) Experiment, STEREO High Energy Telescope (HET). We are grateful to the 
SPDF OMNIWeb interface at {http://omniweb.gsfc.nasa.gov} for the solar and 
interplanetary data. We also thank the providers of NM data used in this study.

\end{acknowledgements}


\bibliographystyle{agu08}
\bibliography{mybib}

\begin{thebibliography}{59}
\providecommand{\natexlab}[1]{#1}
\expandafter\ifx\csname urlstyle\endcsname\relax
  \providecommand{\doi}[1]{doi:\discretionary{}{}{}#1}\else
  \providecommand{\doi}{doi:\discretionary{}{}{}\begingroup
  \urlstyle{rm}\Url}\fi

\bibitem[{\textit{{Adriani} et~al.}(2011)}]{Adriani2011Science}
{Adriani}, O., et~al. (2011), {{PAMELA} measurements of cosmic-ray proton and
  helium spectra}, \textit{Science}, \textit{332}(6025), 69--72,
  \doi{10.1126/science.1199172}.

\bibitem[{\textit{{Adriani} et~al.}(2013)}]{adr13}
{Adriani}, O., et~al. (2013), {Time eependence of the proton flux measured by
  PAMELA during the 2006 July-2009 December solar minimum}, \textit{Astrophys.
  J.}, \textit{765}, 91, \doi{10.1088/0004-637X/765/2/91}.

\bibitem[{\textit{Ahluwalia et~al.}(2010)\textit{Ahluwalia, Lopate, Ygbuhay,
  and Duldig}}]{ahl10}
Ahluwalia, H.~S., C.~Lopate, R.~C. Ygbuhay, and M.~L. Duldig (2010), Galactic
  cosmic ray modulation for sunspot cycle 23, \textit{Adv. Space Res.},
  \textit{46}(7), 934 -- 941,
  \doi{http://dx.doi.org/10.1016/j.asr.2010.04.008}.

\bibitem[{\textit{{Alanko} et~al.}(2003)\textit{{Alanko}, {Usoskin}, {Mursula},
  and {Kovaltsov}}}]{ala03}
{Alanko}, K., I.~G. {Usoskin}, K.~{Mursula}, and G.~A. {Kovaltsov} (2003),
  {Heliospheric modulation strength: effective neutron monitor energy},
  \textit{Adv. Space Res.}, \textit{32}(4), 615--620,
  \doi{10.1016/S0273-1177(03)00348-X}.

\bibitem[{\textit{Ball et~al.}(2005)\textit{Ball, Zhang, Rassoul, and
  Linde}}]{bal05}
Ball, B., M.~Zhang, H.~Rassoul, and T.~Linde (2005), Galactic cosmic-ray
  modulation using a solar minimum {MHD} heliosphere: a stochastic particle
  approach, \textit{Astrophys. J.}, \textit{634}(2), 1116,
  \doi{10.1086/496965}.

\bibitem[{\textit{Burger}(2012)}]{bur12}
Burger, R.~A. (2012), Modeling drift along the heliospheric wavy neutral sheet,
  \textit{Astrophys. J.}, \textit{760}(1), 60,
  \doi{10.1088/0004-637X/760/1/60}.

\bibitem[{\textit{Burger and Potgieter}(1989)}]{bur89}
Burger, R.~A., and M.~S. Potgieter (1989), The calculation of neutral sheet
  drift in two-dimensional cosmic-ray modulation models, \textit{Astrophys.
  J.}, \textit{339}, 501--511, \doi{10.1086/167313}.

\bibitem[{\textit{{B{\"u}sching} and {Potgieter}}(2008)}]{bus08}
{B{\"u}sching}, I., and M.~S. {Potgieter} (2008), {The variability of the
  proton cosmic ray flux on the Sun's way around the galactic center},
  \textit{Adv. Space Res.}, \textit{42}, 504--509,
  \doi{10.1016/j.asr.2007.05.051}.

\bibitem[{\textit{{Caballero-Lopez} and {Moraal}}(2004)}]{Lopez2004JGR}
{Caballero-Lopez}, R.~A., and H.~{Moraal} (2004), {Limitations of the force
  field equation to describe cosmic ray modulation}, \textit{J. Geophys. Res.},
  \textit{109}, A01101, \doi{10.1029/2003JA010098}.

\bibitem[{\textit{{Cliver} et~al.}(2013)\textit{{Cliver}, {Richardson}, and
  {Ling}}}]{Cliver2011SSR}
{Cliver}, E.~W., I.~G. {Richardson}, and A.~G. {Ling} (2013), {Solar Drivers of
  11-yr and Long-Term Cosmic Ray Modulation}, \textit{Space Sci. Rev.},
  \textit{176}, 3--19, \doi{10.1007/s11214-011-9746-3}.

\bibitem[{\textit{{Decker} et~al.}(2012)\textit{{Decker}, {Krimigis}, {Roelof},
  and {Hill}}}]{Decker2012Nature}
{Decker}, R.~B., S.~M. {Krimigis}, E.~C. {Roelof}, and M.~E. {Hill} (2012), {No
  meridional plasma flow in the heliosheath transition region},
  \textit{Nature}, \textit{489}, 124--127, \doi{10.1038/nature11441}.

\bibitem[{\textit{{Effenberger}
  et~al.}(2012{\natexlab{a}})\textit{{Effenberger}, {Fichtner}, {Scherer},
  {Barra}, {Kleimann}, and {Strauss}}}]{eff12apj}
{Effenberger}, F., H.~{Fichtner}, K.~{Scherer}, S.~{Barra}, J.~{Kleimann}, and
  R.~D. {Strauss} (2012{\natexlab{a}}), {A generalized diffusion tensor for
  fully anisotropic diffusion of energetic particles in the heliospheric
  magnetic field}, \textit{Astrophys. J.}, \textit{750}, 108,
  \doi{10.1088/0004-637X/750/2/108}.

\bibitem[{\textit{{Effenberger}
  et~al.}(2012{\natexlab{b}})\textit{{Effenberger}, {Fichtner}, {Scherer}, and
  {B{\"u}sching}}}]{eff12}
{Effenberger}, F., H.~{Fichtner}, K.~{Scherer}, and I.~{B{\"u}sching}
  (2012{\natexlab{b}}), {Anisotropic diffusion of galactic cosmic ray protons
  and their steady-state azimuthal distribution}, \textit{Astrophys. J.},
  \textit{547}, A120, \doi{10.1051/0004-6361/201220203}.

\bibitem[{\textit{{Ferreira} et~al.}(2001)\textit{{Ferreira}, {Potgieter},
  {Burger}, {Heber}, and {Fichtner}}}]{fer01}
{Ferreira}, S.~E.~S., M.~S. {Potgieter}, R.~A. {Burger}, B.~{Heber}, and
  H.~{Fichtner} (2001), {Modulation of Jovian and galactic electrons in the
  heliosphere: 1. Latitudinal transport of a few MeV electrons}, \textit{J.
  Geophys. Res.}, \textit{106}, 24,979--24,988, \doi{10.1029/2001JA000082}.

\bibitem[{\textit{{Giacalone} and {Jokipii}}(1999)}]{gia99}
{Giacalone}, J., and J.~R. {Jokipii} (1999), {The transport of cosmic rays
  across a turbulent magnetic field}, \textit{Astrophys. J.}, \textit{520},
  204--214, \doi{10.1086/307452}.

\bibitem[{\textit{Goring and Nikora}(2002)}]{gor02}
Goring, D., and V.~Nikora (2002), Despiking acoustic doppler velocimeter data,
  \textit{J. Hydrauli. Eng.}, \textit{128}(1), 117--126,
  \doi{10.1061/(ASCE)0733-9429(2002)128:1(117)}.

\bibitem[{\textit{{Heber} et~al.}(2006)\textit{{Heber}, {Fichtner}, and
  {Scherer}}}]{heb06}
{Heber}, B., H.~{Fichtner}, and K.~{Scherer} (2006), {Solar and heliospheric
  modulation of galactic cosmic rays}, \textit{Space Sci. Rev.}, \textit{125},
  81--93, \doi{10.1007/s11214-006-9048-3}.

\bibitem[{\textit{{Heber} et~al.}(2009)\textit{{Heber}, {Kopp}, {Gieseler},
  {M{\"u}ller-Mellin}, {Fichtner}, {Scherer}, {Potgieter}, and
  {Ferreira}}}]{heb09}
{Heber}, B., A.~{Kopp}, J.~{Gieseler}, R.~{M{\"u}ller-Mellin}, H.~{Fichtner},
  K.~{Scherer}, M.~S. {Potgieter}, and S.~E.~S. {Ferreira} (2009), {Modulation
  of galactic cosmic ray protons and electrons during an unusual solar
  minimum}, \textit{Astrophys. J.}, \textit{699}, 1956--1963,
  \doi{10.1088/0004-637X/699/2/1956}.

\bibitem[{\textit{{Herbst} et~al.}(2010)\textit{{Herbst}, {Kopp}, {Heber},
  {Steinhilber}, {Fichtner}, {Scherer}, and {Matthi{\"a}}}}]{herbst10}
{Herbst}, K., A.~{Kopp}, B.~{Heber}, F.~{Steinhilber}, H.~{Fichtner},
  K.~{Scherer}, and D.~{Matthi{\"a}} (2010), {On the importance of the local
  interstellar spectrum for the solar modulation parameter}, \textit{J.
  Geophys. Res.}, \textit{115}, D00I20, \doi{10.1029/2009JD012557}.

\bibitem[{\textit{{Herbst} et~al.}(2012)\textit{{Herbst}, {Heber}, {Kopp},
  {Sternal}, and {Steinhilber}}}]{Herbst2012APJ}
{Herbst}, K., B.~{Heber}, A.~{Kopp}, O.~{Sternal}, and F.~{Steinhilber} (2012),
  {The local interstellar spectrum beyond the heliopause: what can be learned
  from {V}oyager in the inner heliosheath?}, \textit{Astrophys. J.},
  \textit{761}, 17, \doi{10.1088/0004-637X/761/1/17}.

\bibitem[{\textit{{Hitge} and {Burger}}(2010)}]{HitgeABurger10}
{Hitge}, M., and R.~A. {Burger} (2010), {Cosmic ray modulation with a Fisk-type
  heliospheric magnetic field and a latitude-dependent solar wind speed},
  \textit{Adv. Space Res.}, \textit{45}, 18--27,
  \doi{10.1016/j.asr.2009.07.024}.

\bibitem[{\textit{{Jokipii}}(1966)}]{jok66}
{Jokipii}, J.~R. (1966), {Cosmic-ray propagation. I. charged particles in a
  random magnetic field}, \textit{Astrophys. J.}, \textit{146}, 480,
  \doi{10.1086/148912}.

\bibitem[{\textit{{Jokipii} and {Kopriva}}(1979)}]{jok79}
{Jokipii}, J.~R., and D.~A. {Kopriva} (1979), {Effects of particle drift on the
  transport of cosmic rays. III - Numerical models of galactic cosmic-ray
  modulation}, \textit{Astrophys. J.}, \textit{234}, 384--392,
  \doi{10.1086/157506}.

\bibitem[{\textit{{Jokipii} and {K{\'o}ta}}(2000)}]{jok00}
{Jokipii}, J.~R., and J.~{K{\'o}ta} (2000), {Galactic and anomalous cosmic rays
  in the heliosphere}, \textit{Astrophys. Space Sci.}, \textit{274}, 77--96,
  \doi{10.1023/A:1026535603934}.

\bibitem[{\textit{{Jokipii} and {Thomas}}(1981)}]{jok81}
{Jokipii}, J.~R., and B.~{Thomas} (1981), {Effects of drift on the transport of
  cosmic rays. IV - Modulation by a wavy interplanetary current sheet},
  \textit{Astrophys. J.}, \textit{243}, 1115--1122, \doi{10.1086/158675}.

\bibitem[{\textit{{Kopp} et~al.}(2012)\textit{{Kopp}, {B{\"u}sching},
  {Strauss}, and {Potgieter}}}]{kopp12}
{Kopp}, A., I.~{B{\"u}sching}, R.~D. {Strauss}, and M.~S. {Potgieter} (2012),
  {A stochastic differential equation code for multidimensional
  {F}okker-{P}lanck type problems}, \textit{Comput. Phys. Commun.},
  \textit{183}, 530--542, \doi{10.1016/j.cpc.2011.11.014}.

\bibitem[{\textit{{K{\'o}ta}}(2013)}]{kot12}
{K{\'o}ta}, J. (2013), {Theory and modeling of galactic cosmic rays: trends and
  prospects}, \textit{Space. Sci. Rev.}, \textit{176}, 391--403,
  \doi{10.1007/s11214-012-9870-8}.

\bibitem[{\textit{{K\'ota} and {Jokipii}}(1983)}]{kot83}
{K\'ota}, J., and J.~R. {Jokipii} (1983), {Effects of drift on the transport of
  cosmic rays. VI - a three-dimensional model including diffusion},
  \textit{Astrophys. J.}, \textit{265}, 573--581, \doi{10.1086/160701}.

\bibitem[{\textit{Manuel et~al.}(2011)\textit{Manuel, Ferreira, Potgieter,
  Strauss, and Engelbrecht}}]{man11}
Manuel, R., S.~E.~S. Ferreira, M.~S. Potgieter, R.~D. Strauss, and N.~E.
  Engelbrecht (2011), Time-dependent cosmic ray modulation, \textit{Adv. Space
  Res.}, \textit{47}(9), 1529 -- 1537, \doi{10.1016/j.asr.2010.12.007}.

\bibitem[{\textit{{Matthaeus} et~al.}(2003)\textit{{Matthaeus}, {Qin},
  {Bieber}, and {Zank}}}]{MatthaeusEA03}
{Matthaeus}, W.~H., G.~{Qin}, J.~W. {Bieber}, and G.~P. {Zank} (2003),
  {Nonlinear collisionless perpendicular diffusion of charged particles},
  \textit{Astrophys. J.}, \textit{590}, L53--L56, \doi{10.1086/376613}.

\bibitem[{\textit{McDonald}(1998)}]{McDonald98}
McDonald, F.~B. (1998), Cosmic-ray modulation in the heliosphere--a
  phenomenological study, \textit{Space Sci. Rev.}, \textit{83}, 33--50,
  \doi{10.1023/A:1005052908493}.

\bibitem[{\textit{{Moraal}}(2013)}]{Moraal2013SSR}
{Moraal}, H. (2013), {Cosmic-Ray Modulation Equations}, \textit{Space Sci.
  Rev.}, \textit{176}, 299--319, \doi{10.1007/s11214-011-9819-3}.

\bibitem[{\textit{Nymmik et~al.}(1992)\textit{Nymmik, Panasyuk, Pervaja, and
  Suslov}}]{nymmik92}
Nymmik, R.~A., M.~I. Panasyuk, T.~I. Pervaja, and A.~A. Suslov (1992), A model
  of galactic cosmic ray fluxes, \textit{Int. J. Rad. Appl. Instrum. Part D.
  Nucl. Track. Radiat. Meas.}, \textit{20}(3), 427--429,
  \doi{http://dx.doi.org/10.1016/1359-0189(92)90028-T}.

\bibitem[{\textit{{Parker}}(1965)}]{par65}
{Parker}, E.~N. (1965), {The passage of energetic charged particles through
  interplanetary space}, \textit{Planet. Space Sci.}, \textit{13}, 9--49,
  \doi{10.1016/0032-0633(65)90131-5}.

\bibitem[{\textit{Pei et~al.}(2010)\textit{Pei, Bieber, Burger, and
  Clem}}]{pei10}
Pei, C., J.~W. Bieber, R.~A. Burger, and J.~Clem (2010), A general
  time-dependent stochastic method for solving parker's transport equation in
  spherical coordinates, \textit{J. Geophys. Res.}, \textit{115}(A12), A12,107.

\bibitem[{\textit{{Pogorelov} et~al.}(2009)\textit{{Pogorelov}, {Heerikhuisen},
  {Zank}, and {Borovikov}}}]{Pogorelov2009SSR}
{Pogorelov}, N.~V., J.~{Heerikhuisen}, G.~P. {Zank}, and S.~N. {Borovikov}
  (2009), {Influence of the interstellar magnetic field and neutrals on the
  shape of the outer heliosphere}, \textit{Space Sci. Rev.}, \textit{143},
  31--42, \doi{10.1007/s11214-008-9429-x}.

\bibitem[{\textit{{Potgieter}}(2013)}]{pot13}
{Potgieter}, M. (2013), {Solar modulation of cosmic rays}, \textit{Living Rev.
  Solar Phys.}, \textit{10}, 3, \doi{10.12942/lrsp-2013-3}.

\bibitem[{\textit{Potgieter}(1998)}]{pot98}
Potgieter, M.~S. (1998), The modulation of galactic cosmic rays in the
  heliosphere: theory and models, \textit{Space Sci. Rev.}, \textit{83}(1),
  147--158, \doi{10.1023/A:1005014722123}.

\bibitem[{\textit{{Potgieter} et~al.}(2014)\textit{{Potgieter}, {Vos},
  {Boezio}, {De Simone}, {Di Felice}, and {Formato}}}]{Potgieter2014SP}
{Potgieter}, M.~S., E.~E. {Vos}, M.~{Boezio}, N.~{De Simone}, V.~{Di Felice},
  and V.~{Formato} (2014), {Modulation of Galactic Protons in the Heliosphere
  During the Unusual Solar Minimum of 2006 to 2009}, \textit{Solar Phys.},
  \textit{289}, 391--406, \doi{10.1007/s11207-013-0324-6}.

\bibitem[{\textit{{Qin}}(2002)}]{qin02}
{Qin}, G. (2002), {Charged particle transport in magnetic field turbulence and
  study of trim simulation and SSX experiment}, Ph.D. thesis, University of
  Delaware.

\bibitem[{\textit{{Qin}}(2007)}]{qin07}
{Qin}, G. (2007), {Nonlinear parallel diffusion of charged particles: extension
  to the nonlinear guiding center theory}, \textit{Astrophys. J.},
  \textit{656}, 217--221, \doi{10.1086/510510}.

\bibitem[{\textit{{Qin} et~al.}(2004)\textit{{Qin}, {Zhang}, {Dwyer}, and
  {Rassoul}}}]{QinEA04}
{Qin}, G., M.~{Zhang}, J.~R. {Dwyer}, and H.~K. {Rassoul} (2004),
  {Interplanetary transport mechanisms of solar energetic particles},
  \textit{Astrophys. J.}, \textit{609}, 1076--1081, \doi{10.1086/421101}.

\bibitem[{\textit{{Qin} et~al.}(2005)\textit{{Qin}, {Zhang}, {Dwyer},
  {Rassoul}, and {Mason}}}]{qin05}
{Qin}, G., M.~{Zhang}, J.~R. {Dwyer}, H.~K. {Rassoul}, and G.~M. {Mason}
  (2005), {The model dependence of solar energetic particle mean free paths
  under weak scattering}, \textit{Astrophys. J.}, \textit{627}, 562--566,
  \doi{10.1086/430136}.

\bibitem[{\textit{Qin et~al.}(2012)\textit{Qin, Zhao, and Chen}}]{qin12}
Qin, G., L.-L. Zhao, and H.-C. Chen (2012), Despiking of spacecraft energetic
  proton flux to study galactic cosmic-ray modulation, \textit{Astrophys. J.},
  \textit{752}(2), 138, \doi{10.1088/0004-637X/752/2/138}.

\bibitem[{\textit{{Reinecke} and {Potgieter}}(1993)}]{rein93}
{Reinecke}, J.~P.~L., and M.~S. {Potgieter} (1993), {An explanation for the
  observed intersection of cosmic-ray spectra for consecutive solar minimum
  periods}, \textit{Int. Cosmic Ray Conf.}, \textit{3}, 597.

\bibitem[{\textit{{Scherer} and {Fahr}}(2003)}]{Scherer2003GRL}
{Scherer}, K., and H.~J. {Fahr} (2003), {Solar cycle induced variations of the
  outer heliospheric structures}, \textit{Geophys. Res. Lett.}, \textit{30}(2),
  1045, \doi{10.1029/2002GL016073}.

\bibitem[{\textit{{Scherer} et~al.}(2004)\textit{{Scherer}, {Fahr}, {Fichtner},
  and {Heber}}}]{sch04}
{Scherer}, K., H.-J. {Fahr}, H.~{Fichtner}, and B.~{Heber} (2004), {Long-term
  modulation of cosmic rays in the heliosphere and its influence at earth},
  \textit{Solar Phys.}, \textit{224}, 305--316,
  \doi{10.1007/s11207-005-5687-x}.

\bibitem[{\textit{{Scherer} et~al.}(2011)\textit{{Scherer}, {Fichtner},
  {Strauss}, {Ferreira}, {Potgieter}, and {Fahr}}}]{Scherer2011APJ}
{Scherer}, K., H.~{Fichtner}, R.~D. {Strauss}, S.~E.~S. {Ferreira}, M.~S.
  {Potgieter}, and H.-J. {Fahr} (2011), {On cosmic ray modulation beyond the
  heliopause: where is the modulation boundary?}, \textit{Astrophys. J.},
  \textit{735}, 128, \doi{10.1088/0004-637X/735/2/128}.

\bibitem[{\textit{{Shalchi} and {B{\"u}sching}}(2010)}]{sha10}
{Shalchi}, A., and I.~{B{\"u}sching} (2010), {Influence of turbulence
  dissipation effects on the propagation of low-energy cosmic rays in the
  galaxy}, \textit{Astrophys. J.}, \textit{725}, 2110--2116,
  \doi{10.1088/0004-637X/725/2/2110}.

\bibitem[{\textit{{Shalchi} et~al.}(2004)\textit{{Shalchi}, {Bieber}, and
  {Matthaeus}}}]{ShalchiEA04}
{Shalchi}, A., J.~W. {Bieber}, and W.~H. {Matthaeus} (2004), {Analytic forms of
  the perpendicular diffusion coefficient in magnetostatic turbulence},
  \textit{Astrophys. J.}, \textit{604}, 675--686, \doi{10.1086/382128}.

\bibitem[{\textit{Strauss and Potgieter}(2014)}]{Strauss2014ASR}
Strauss, R.~D., and M.~S. Potgieter (2014), Where does the heliospheric
  modulation of galactic cosmic rays start?, \textit{Adv. Space Res.},
  \doi{http://dx.doi.org/10.1016/j.asr.2014.01.004}.

\bibitem[{\textit{{Strauss} et~al.}(2011)\textit{{Strauss}, {Potgieter},
  {Kopp}, and {B{\"u}sching}}}]{Strauss2011JGR}
{Strauss}, R.~D., M.~S. {Potgieter}, A.~{Kopp}, and I.~{B{\"u}sching} (2011),
  {On the propagation times and energy losses of cosmic rays in the
  heliosphere}, \textit{J. Geophys. Res.}, \textit{116}, A12105,
  \doi{10.1029/2011JA016831}.

\bibitem[{\textit{{Strauss} et~al.}(2012)\textit{{Strauss}, {Potgieter},
  {B{\"u}sching}, and {Kopp}}}]{stra12}
{Strauss}, R.~D., M.~S. {Potgieter}, I.~{B{\"u}sching}, and A.~{Kopp} (2012),
  {Modelling heliospheric current sheet drift in stochastic cosmic ray
  transport models}, \textit{Astrophys. Space Sci.}, \textit{339}, 223--236,
  \doi{10.1007/s10509-012-1003-z}.

\bibitem[{\textit{{Strauss} et~al.}(2013)\textit{{Strauss}, {Potgieter},
  {Ferreira}, {Fichtner}, and {Scherer}}}]{Strauss2013APJL}
{Strauss}, R.~D., M.~S. {Potgieter}, S.~E.~S. {Ferreira}, H.~{Fichtner}, and
  K.~{Scherer} (2013), {Cosmic ray modulation beyond the heliopause: a hybrid
  modeling approach}, \textit{Astrophys. J. Lett.}, \textit{765}, L18,
  \doi{10.1088/2041-8205/765/1/L18}.

\bibitem[{\textit{{Webber} et~al.}(2013)\textit{{Webber}, {Higbie}, and
  {McDonald}}}]{Webber2013}
{Webber}, W.~R., P.~R. {Higbie}, and F.~B. {McDonald} (2013), {The unfolding of
  the spectra of low energy galactic cosmic ray {H} and {H}e nuclei as the
  {V}oyager 1 spacecraft exits the region of heliospheric modulation},
  \textit{ArXiv e-prints}.

\bibitem[{\textit{{Zank} and {M{\"u}ller}}(2003)}]{Zank2003JGR}
{Zank}, G.~P., and H.-R. {M{\"u}ller} (2003), {The dynamical heliosphere},
  \textit{J. Geophys. Res.}, \textit{108}, 1240, \doi{10.1029/2002JA009689}.

\bibitem[{\textit{{Zank} et~al.}(2004)\textit{{Zank}, {Li}, {Florinski},
  {Matthaeus}, {Webb}, and {Le Roux}}}]{ZankEA04}
{Zank}, G.~P., G.~{Li}, V.~{Florinski}, W.~H. {Matthaeus}, G.~M. {Webb}, and
  J.~A. {Le Roux} (2004), {Perpendicular diffusion coefficient for charged
  particles of arbitrary energy}, \textit{J. Geophys. Res.}, \textit{109},
  A04107, \doi{10.1029/2003JA010301}.

\bibitem[{\textit{{Zhang}}(1999)}]{zha99}
{Zhang}, M. (1999), {A markov stochastic process theory of cosmic-ray
  modulation}, \textit{Astrophys. J.}, \textit{513}, 409--420,
  \doi{10.1086/306857}.

\bibitem[{\textit{Zhao and Qin}(2013)}]{zhao12}
Zhao, L.-L., and G.~Qin (2013), An observation-based {GCR} model of heavy
  nuclei: measurements from {CRIS} onboard {ACE} spacecraft, \textit{J.
  Geophys. Res.}, pp. 1837--1848, \doi{10.1002/jgra.50235}.

\end{thebibliography}


\end{article}


\begin{table}[p]
\caption{Values of parameters used in the simulations for the last
three solar minima. \label{tbl-pms}}
\begin{flushleft}
\begin{tabular*}{\textwidth}{@{\extracolsep{\fill}}cccc}
\tableline
Parameter & $P_{21/22}$ & $P_{22/23}$ & $P_{23/24}$ \\
\tableline
$A$    & $<0$            &  $>0$        & $<0$                      \\
$V_{sw}$  & $442$ km/s   & $416$ km/s   & $360$ km/s    \\
$B_e$  & $5.5$ nT           & $4.9$ nT           & $3.9$ nT         \\
$\alpha$    & $4.3^\circ$       & $4.3^\circ$         & $6.3^\circ$ \\
$a$         &  $0.03$              & $0.03$          & $0.02$     \\
$b$          & $0.02$      & $0.02$            & $0.01$ \\
$d$          & $0.5$      & $0.5$            & $1$ \\
\tableline
\end{tabular*}
\end{flushleft}
\end{table}
\clearpage
\begin{table}[p]
  \caption{Neutron Monitors(NMs) and parameters used in the comparison for the last
 three solar minima. \label{tbl-nms}}
 \begin{flushleft}
 \begin{tabular*}{\textwidth}{@{\extracolsep{\fill}}cccc}
 \tableline
  NM & $P_c$ (GV) & $E_{\text{eff}}$ (GeV) & $K_{\text{NM}}$ 
(m${}^{-2}$sr${}^{-1}$GeV${}^{-1}$)\\
 \tableline
 Apatity         & $0.65$        & $6.50$ & $3.16\times 10^{-5}$\\
 Oulu            & $0.80$        & $6.54$ & $3.53\times 10^{-5}$\\
 Yakutsk         & $1.65$        & $6.87$ & $3.43\times 10^{-5}$\\
 Moscow          & $2.43$        & $7.41$ & $2.07\times 10^{-5}$\\
 Novosibirsk     & $2.91$        & $7.89$ & $3.43\times 10^{-5}$\\
 Lomnicky Stit   & $3.98$        & $9.46$ & $8.72\times 10^{-5}$\\
 Jungfraujoch    & $4.49$        & $10.47$& $2.05\times 10^{-5}$\\
 Hermanus        & $4.58$        & $10.67$& $2.63\times 10^{-5}$\\
 Rome            & $6.32$        & $15.59$& $1.44\times 10^{-5}$\\
 Tbilisi         & $6.73$        & $16.99$& $6.78\times 10^{-5}$\\
 Potchefstroom   & $7.00$        & $17.96$& $2.69\times 10^{-5}$\\
\tableline
\end{tabular*}
\end{flushleft}
\end{table}
\clearpage


%
%
%
%
%
%


\clearpage

\clearpage

\end{document}